\documentclass{article}
\usepackage{spconf,amsmath,amssymb,graphicx,bm,algorithm,algpseudocode,mathtools,xcolor,siunitx}
\graphicspath{{figures/}}

\providecommand{\norm}[1]{\left\lVert#1\right\rVert}

\title{Optimal Measurement Configuration in Computational \\ Diffractive Imaging}

\name{Evan Widloski \qquad Ulas Kamaci \qquad Farzad Kamalabadi}
\address{
Department of Electrical and Computer Engineering and Coordinated Science Laboratory, \\
University of Illinois at Urbana-Champaign, Urbana, IL 61801, USA
}

\begin{document}
\maketitle

\begin{abstract}

Diffractive lenses have recently been applied to the domain of multispectral
imaging in the X-ray and UV regimes where they can achieve very high resolution
as compared to reflective and refractive optics. Conventionally, spectral
components are reconstructed by taking measurements at the focal planes.
However, the reconstruction quality can be improved by optimizing the
measurement configuration. In this work, we adapt a sequential backward
selection algorithm to search for a configuration which minimizes expected
reconstruction error. By approximating the forward system as a circular
convolution and making assumptions on the source and noise, we greatly reduce
the complexity of the algorithm. Numerical results show that the configuration
found by the algorithm significantly improves the reconstruction performance
compared to a standard configuration.

\end{abstract}


\section{Introduction}
\label{sec:intro}
Spectral imaging is the formation of images at different wavelengths in the
electromagnetic spectrum. With images usually taken in the visible, X-ray,
ultraviolet (UV), or infrared bands, it has applications in medicine,
geographic surveying, astronomy, and solar physics \cite{shaw2003spectral},
\cite{garini2006spectral}. In spectral imaging, a polychromatic source must be first separated
into its spectral components before being captured. There are a number of ways
to achieve this, but a common method is to use a set of configurable optical
filters. For example, the spectral imager on the Solar Dynamics Observatory uses
a rotating drum of optical filters to selectively pass light of specific
wavelengths of interest \cite{sdo}.

\begin{figure}[htb]

\begin{minipage}[b]{0.48\linewidth}
  \centering
  \centerline{\includegraphics[width=4.0cm]{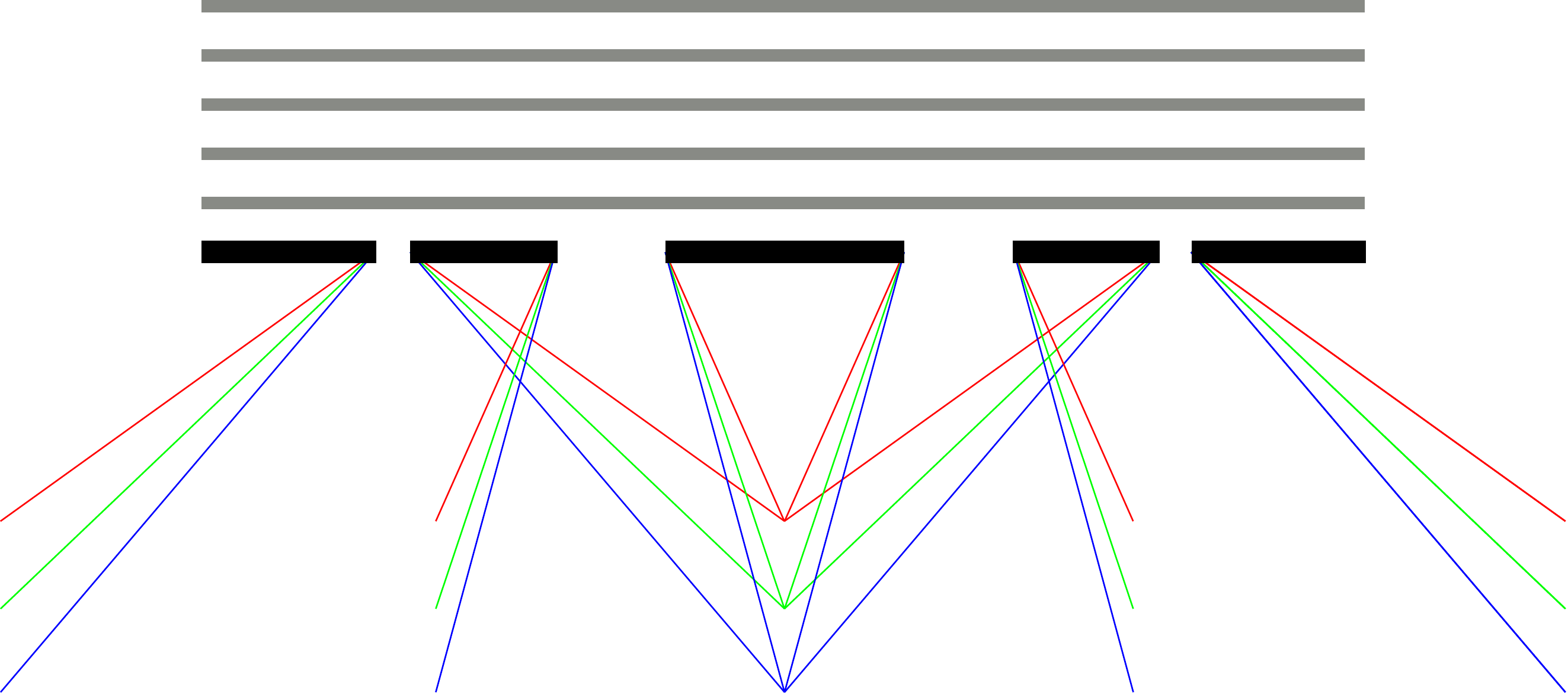}}
  \centerline{(a)}\medskip
\end{minipage}
\hfill
\begin{minipage}[b]{0.24\linewidth}
  \centering
  \centerline{\includegraphics[width=2.0cm]{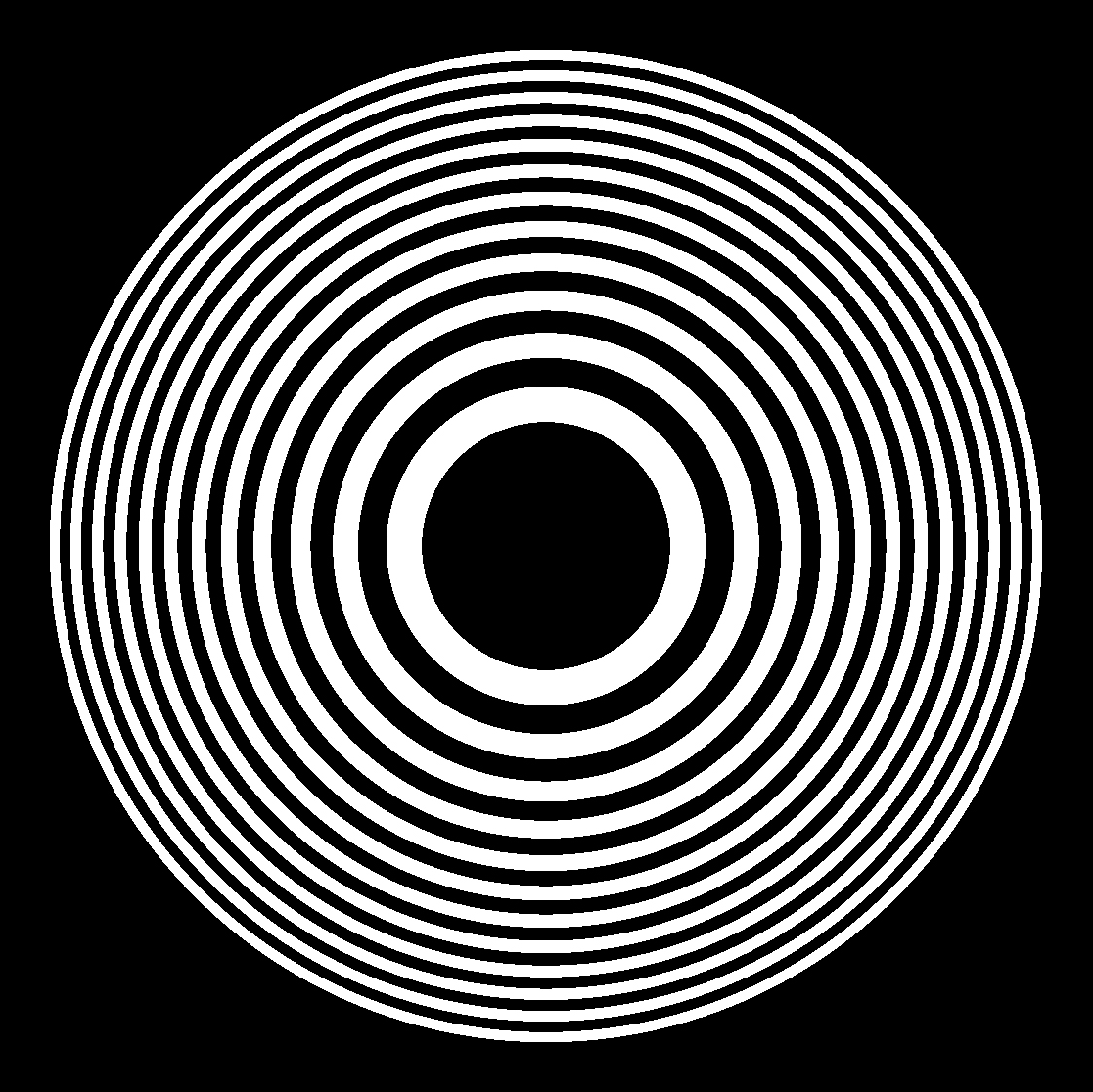}}
  \centerline{(b)}\medskip
\end{minipage}
\hfill
\begin{minipage}[b]{0.24\linewidth}
  \centering
  \centerline{\includegraphics[width=2.0cm]{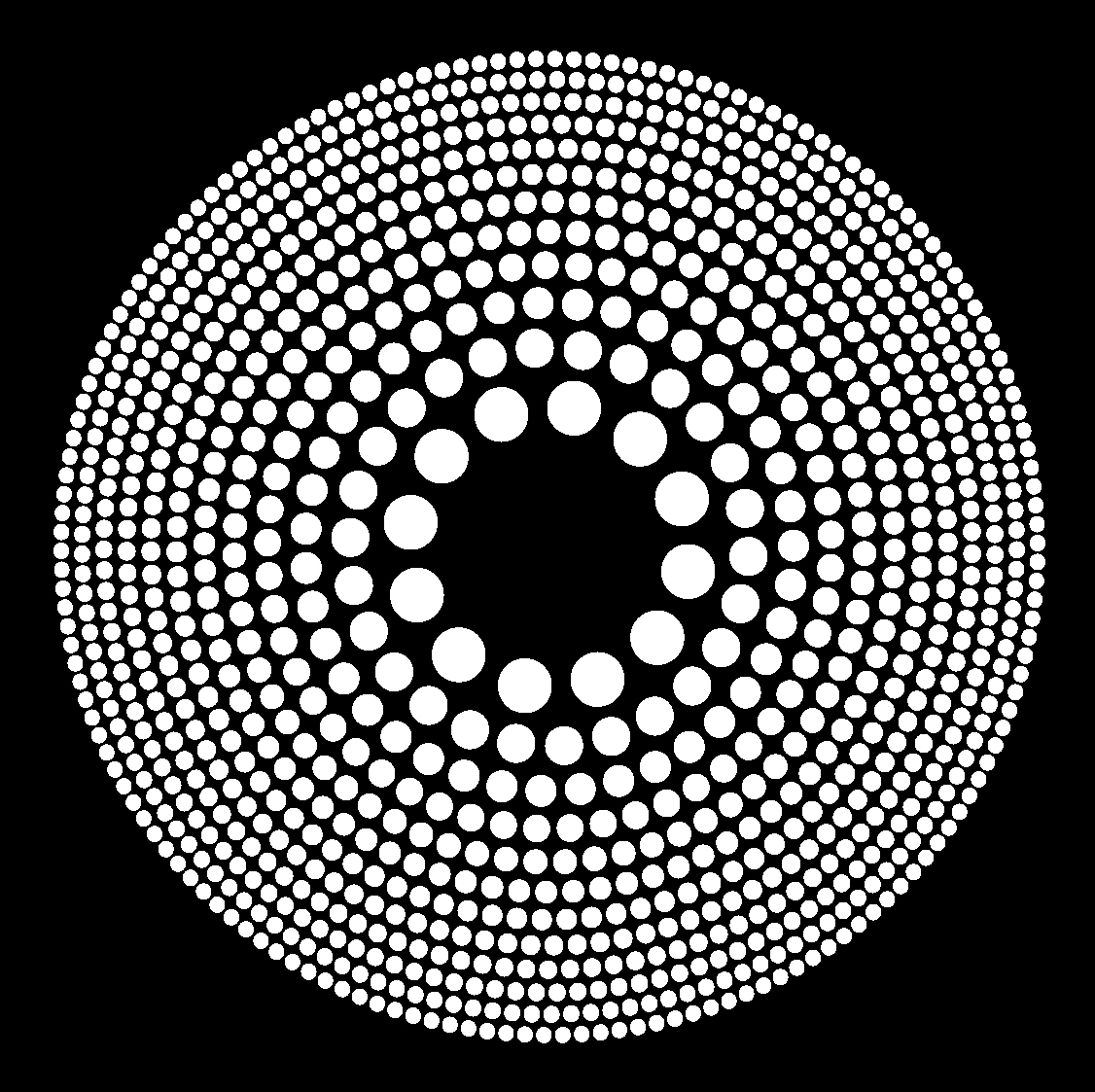}}
  \centerline{(c)}\medskip
\end{minipage}
\caption{(a) diffraction of a polychromatic wave through a diffractive lens (b) Fresnel zone
plate (c) photon sieve \cite{kipp2001sharper}}
\label{fig:diff_lens}
\end{figure}

A new approach is to use a diffractive lens to perform spectral imaging
\cite{oktem2014icip}.
Diffractive lenses are often preferred in the UV or X-ray regimes because
manufacturing tolerances at these wavelengths can be more relaxed than
reflective optics and still obtain a similar resolution \cite{davila2011high}.
Since diffractive optics can be manufactured using a photolithographic process,
they can be produced at a higher precision compared to the grinding process used
to produce conventional reflective optics. Moreover, refractive optics are
unsuitable for UV or X-ray imaging because glass is opaque at these wavelengths.
Figures \ref{fig:diff_lens}(b) and \ref{fig:diff_lens}(c) are two examples of a
pattern that can be etched into silicon wafer to produce a diffractive lens.

Diffractive lenses have the property that the angle at which light exits the
lens is determined by the light's wavelength, which gives them a wavelength
dependent focal length, as shown in Figure \ref{fig:diff_lens}(a)
\cite{attwood1999}.

\begin{figure}[htb]
  \begin{minipage}[b]{1\linewidth}
    \centering
    \centerline{\includegraphics[width=8.5cm]{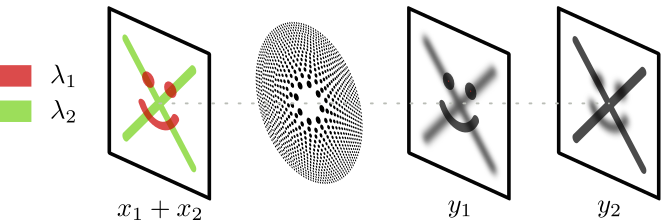}}
  \end{minipage}
  \caption{Imaging a scene with emissions at wavelengths $\lambda_1$ and
  $\lambda_2$. Measurements $y_1$ and $y_2$ are taken at two positions where one wavelength
  is in focus and the other is out of focus.}
  \label{fig:pssi_drawing}
\end{figure}

Measurements at the focal plane of each spectral component comprise of a sum of a
focused image of one component and blurred images of all other
components, as shown in Figure \ref{fig:pssi_drawing}.
An inverse problem consisting of disentangling and deblurring of measurements
must be solved in order to recover the original source components
\cite{oktem2014icip}. However, this focal plane measurement configuration leads
to suboptimal reconstructions, especially when spectral components are close in
wavelength. Therefore, it is desired to determine the optimal measurement
configuration before acquiring the data.

Finding the optimal measurement configuration can be seen as a \emph{sensor
placement problem}, which lies under the broader class of problems known as
\emph{subset selection}. Subset selection applies to many domains, such as array
optimization for atmospheric imaging \cite{sharif}, \cite{wang2017near},
magnetic resonance imaging \cite{gao2000optimal}, and detection problems
\cite{yu1997sampling}. Methods like genetic algorithms, convex optimization
\cite{joshi2008sensor}, and hill climbing \cite{reeves1999sequential} with many
selection criteria have been developed to solve such problems.

However, most of these methods solve the problem of
single-sensor/single-measurement systems where the placement of one sensor
contributes a single row to the observation matrix. In contrast, many imaging
systems are single-sensor/multiple-measurement (like our problem), where each sensor placed (measurement plane) contributes multiple rows to the observation matrix (one row
per detector pixel). Single-sensor/single-measurement algorithms have been
extended to the multiple measurement case, known as \emph{clustering}
algorithms. Examples include \emph{clustered sequential backward selection} (CSBS)
\cite{sharif}, \emph{clustered FrameSense} (CFS) \cite{ranieri2014near}, \emph{clustered maximum projection on minimum eigenspace} (CMPME) \cite{wang2018sampling}.

In this paper we adapt CSBS to the diffractive imaging problem to automatically
determine a measurement configuration from a set of candidate plane locations,
which minimizes expected reconstruction error. Furthermore, we exploit
structures in the imaging model to make the algorithm computationally feasible
for large images.

\section{Forward Model and Statistical Formulation}
\label{sec:format}
In this section, we mathematically model a diffractive imaging system and
describe the process of recovering the spectral components. Consider a
polychromatic source that has $S$ spectral components $\bm{x}_1, \dots,
\bm{x}_S \in \mathbb R^{N_1\times N_1}$. Using a moving detector, we make $M$ measurements $\bm{y}_1, \dots,
\bm{y}_M \in \mathbb R^{N_2\times N_2}$ at distances $d_1, \dots, d_M$ from the lens. We allow for repeated
measurements at the same plane for a more flexible model that can take into
account non equal exposure times.
Due to linearity, each measurement is a superposition of blurred versions of the
$S$ sources. More formally,

\begin{equation}
\bm{y}_m = \sum_{s=1}^S \bm{a}_{m,s} \ast \bm{x}_s + \bm{n}_m
\label{eq:fwd_model}
\end{equation}

where $\bm{a}_{m,s} \in \mathbb R^{P\times P}$ is a blurring kernel known as a
\emph{point spread function} (PSF), $\ast$ is a 2D convolution, and $\bm{n}_m
\in \mathbb R^{N_2 \times N_2}$ is additive measurement noise. Each PSF depends
on the associated source wavelength and measurement location together with the
diffractive lens parameters and can be computed efficiently
\cite{ayazgok2020efficient}.

Since convolution is a linear operation, we can rewrite the above equation as a
linear system

\begin{equation}
  \underbrace{
    \begin{bmatrix}\bm{y}_1 \\ \vdots \\ \bm{y}_M\end{bmatrix}
  }_{\bm{y}}
  =
  \underbrace{
    \begin{bmatrix}
      \bm{A}_{1, 1} & \hdots & \bm{A}_{1, S} \\
      \vdots & & \vdots \\
      \bm{A}_{M, 1} & \hdots & \bm{A}_{M, S}
    \end{bmatrix}
  }_{\bm{A}_{\bm{d}}}
  \underbrace{
    \begin{bmatrix}\bm{x}_1 \\ \vdots \\ \bm{x}_S\end{bmatrix}
  }_{\bm{x}}
  +
  \underbrace{
    \begin{bmatrix}\bm{n}_1 \\ \vdots \\ \bm{n}_M\end{bmatrix}
  }_{\bm{n}}
\label{eq:fourier_mtx}
\end{equation}


where $\bm{y}_m \in \mathbb R^{N_2^2\times 1}$, $\bm{x}_s \in \mathbb R^{N_1^2\times 1}$ and $\bm{n}_m \in \mathbb R^{N_2^2\times 1}$ have been flattened from their
original 2D shape, and each $\bm{A}_{m,s} \in \mathbb R^{N_2^2\times N_1^2}$ is a block-toeplitz
matrix with toeplitz blocks formed from 2D convolution with PSF $\bm{a}_{m,s}$.
We will refer to the matrix containing all $\bm{A}_{m,s}$
generated by measurements taken at $\bm{d} = \{d_1, \dots, d_M\}$ as
$\bm{A}_{\bm{d}}$.

The problem of where to take measurements $\bm{y}_1, \dots, \bm{y}_M$ has not
been addressed and affects the reconstruction quality.  In order to compare the
impact of different measurement configurations on the reconstruction, it is
necessary to define some cost for the measurement matrix $\bm{A}_{\bm{d}}$.  A
common cost metric is the expected reconstruction error, or expected \emph{sum
of squared errors} (SSE). However, we must have some strategy for the recovery
of $\bm{x}$ to get reconstruction error and we must make statistical assumptions
about $\bm{x}$.\emph{ Maximum a posteriori} (MAP) estimation is one such
strategy.

We assume the original spectral components and noise are distributed according to a
normal distribution such that $\bm{x} \sim \mathcal{N}(\bm{x}_{0}, \bm{\Sigma}_{\bm{x}})$ and
$\bm{n} \sim \mathcal{N}(0, \bm{\Sigma}_{\bm{n}})$.  The MAP estimate is then

$$
\begin{aligned}
  \bm{x}_{MAP} &= \arg \max_{\bm{x} \in \mathbb{C}^n} p(\bm{x} | \bm{y})
  = \arg \max_{\bm{x}} p(\bm{y}|\bm{x}) p(\bm{x})\\
  &= \arg \min_{\bm{x}} \left[  - \log(p(\bm{y}|\bm{x})) - \log
  p(\bm{x})\right] \\
  &= \bm{x}_0 + \left( \bm{A}_{\bm{d}}^H\bm{\Sigma}_{\bm{n}}^{-1} \bm{A}_{\bm{d}} +
    \bm{\Sigma}_{\bm{x}}^{-1}\right)^{-1}
  \cdot  \bm{A}_{\bm{d}}^H \bm{\Sigma}_{\bm{n}}^{-1} (\bm{y} - \bm{A}_{\bm{d}} \bm{x}_0)
\end{aligned}
$$
The reconstruction error is defined as $\bm{e} = \bm{x} - \bm{x_{\text{MAP}}}$,
and the expected sum of squared error cost is $E[\norm{\bm{e}}_2^2]$. This
expression can be rewritten in terms of the error covariance:

\begin{align*}
\mathbb{E}[\norm{\bm{e}}_2^2] =  \mathbb{E}[\bm{e}^H\bm{e}] & = \mathbb{E}[\text{tr}(\bm{e}^H\bm{e})] = \mathbb{E}[\text{tr}(\bm{e}\bm{e}^H)] \\
& = \text{tr}(\mathbb{E}[\bm{e}\bm{e}^H]) = \text{tr}(\bm{\Sigma}_{\bm{e}})
\end{align*}

where the error covariance matrix is defined as $\bm{\Sigma}_{\bm{e}} =
E[\bm{e}\bm{e}^H]$ and has the closed form expression:
\begin{equation}
\bm{\Sigma}_{\bm{e}} = \left( \bm{A}_{\bm{d}}^H\bm{\Sigma}_{\bm{n}}^{-1} \bm{A}_{\bm{d}} +
    \bm{\Sigma}_{\bm{x}}^{-1}\right)^{-1}
\end{equation}

Combining the above equations, we can now write a cost metric which lets us evalute
the expected reconstruction error for a particular measurement configuration $\bm{d}$:
\begin{equation}
\text{Cost}(\bm{d}) = \mathbb{E}[\norm{\bm{e}}_2^2] = \text{tr}\left(\left( \bm{A}_{\bm{d}}^H\bm{\Sigma}_{\bm{n}}^{-1} \bm{A}_{\bm{d}} +
    \bm{\Sigma}_{\bm{x}}^{-1}\right)^{-1}\right)
\end{equation}

\section{Measurement Selection Algorithm}

With a method of evaluating the effect a particular configuration $\bm{d}$ has
on reconstruction error, we can begin considering which configurations are best
suited for minimizing error.  For example, if we are provided with a set of $C$
candidate measurement locations, we may wish to find a subset of size $M$ which
minimizes reconstruction error. This is known as a \emph{subset selection} problem.
One might think to simply search over all possible measurement configurations
of size $M$, but this exhaustive search requires $\binom{C}{M}$ evalutions of
cost, growing on the order of $O(C^M)$.

CSBS is an alternative method which is more computationally feasible, where one
measurement location is eliminated from $\bm{d}$ in each iteration until only
$M$ locations remain. As reconstruction error generally increases as the number
of measurements decreases, CSBS selects for elimination the measurement that
incurs the smallest increase in cost in each iteration.

\begin{algorithm}
  \begin{algorithmic}
    \caption{CSBS Algorithm}
    \State $\bm{d} = \{d_1, \dots, d_C\}$
    \Repeat
      \State $d' = \arg \min_{d \in \bm{d}} \text{Cost}(\bm{d} \backslash \{d\}))$
      \State $\bm{d} = \bm{d} \backslash \{d'\}$
    \Until{$|\bm{d}| = M$}
  \end{algorithmic}
\end{algorithm}

Unlike an exhaustive search, the complexity of CSBS is not combinatorial.  As the
size of $\bm{d}$ shrinks with each iteration, the number of cost evaluations for each
minimization step decreases.  The total number of cost evaluations is
\vspace{-0.1 in}
$$
\begin{aligned}
  \sum_{|\bm{d}| = M}^C |\bm{d}|
  &= O(C^2 - M^2)
\end{aligned}
$$



\vspace{-0.1 in}
\section{Fast Implementation}
While the SSE cost applies to any general linear system, we can augment the
complexity reduction achieved by CSBS by making assumptions about the structure
of $\bm{A}_{\bm{d}}$, $\bm{\Sigma}_{\bm{n}}$ and $\bm{\Sigma}_{\bm{x}}$.
Specifically, if we assume the blocks of these matrices are block-circulant with
circulant blocks (BCCB), then they can be diagonalized by the 2D DFT matrix
where operations involving multiplications and inversions are much faster.

For $\bm{A}_{\bm{d}}$ this means that each block $\bm A_{m,s}$ is BCCB and
corresponds to circular convolution with the kernel $\bm a_{m,s}$. For $\bm
\Sigma_n$, we assume independent noise among image pixels and measurement planes
with variance $(1/\lambda)$, so $\bm \Sigma_n=(1/\lambda) \bm I$ where $\bm I$
represents the identity matrix. For $\bm \Sigma_x$, each of its blocks being
BCCB means that the covariance among image pixels are represented by 2D
convolution kernels.

Assuming $N \times N$ images, each $N^2 \times N^2$ block $\bm A_{m,s}$ of $\bm
{A_d}$ can be decomposed as $\bm A_{m,s} = \bm F^{-1} \widetilde{\bm A}_{m,s}
\bm F$ where $\widetilde{\bm A}_{m,s}$ is the diagonal matrix consisting of the
2D DFT of $\bm a_{m,s}$, and $\bm F$ is the 2D DFT matrix. This yields
\vspace{-0.1 in}
\begin{equation}
  \bm {A_d}=
  \underbrace{
    \begin{bmatrix}
      \bm F^{-1} \hspace{-0.2 in}& \hspace{-0.2 in} & \hspace{-0.20 in} \vspace{-0.1 in}\text{\Large 0} \\
        \vspace{-0.05 in}&\hspace{-0.20 in} \ddots &  \\
      \text{\Large 0} \hspace{-0.17 in}& \hspace{-0.2 in} & \hspace{-0.15 in} \bm F^{-1}
    \end{bmatrix}
  }_{\widetilde{\bm F}^{-1}}
  \underbrace{
    \begin{bmatrix}
      \widetilde{\bm A}_{1,1} \hspace{-0.1 in}&\hspace{-0.1 in} \hdots & \hspace{-0.1 in} \widetilde{\bm A}_{1,S} \vspace{-0.02 in}\\
      \vspace{-0.02 in}\vdots &\hspace{-0.1 in} \ddots& \vdots \\
      \widetilde{\bm A}_{M,1} \hspace{-0.1 in}&\hspace{-0.1 in} \hdots & \hspace{-0.1 in} \widetilde{\bm A}_{M,S} \\
    \end{bmatrix}
  }_{\widetilde{\bm A}_{\bm d}}
  \underbrace{
    \begin{bmatrix}
      \bm F \hspace{-0.1 in}& & \hspace{-0.18 in} \vspace{-0.1 in}\text{\Large 0} \\
        \vspace{-0.05 in}& \hspace{-0.12 in}\ddots &  \\
  \hspace{0.05 in} \text{\Large 0} & & \hspace{-0.12 in} \bm F
    \end{bmatrix}
  }_{\widetilde{\bm F}}
\label{eq:fourier_mtx}
\end{equation}
\vspace{-0.1 in}

so, we have $\bm{A_d} = \widetilde{\bm F}^{-1} \widetilde{\bm A}_{\bm d}
\widetilde{\bm F}$, from which we get $\bm{A_d}^H \bm{A_d} = \widetilde{\bm
F}^{-1} \widetilde{\bm A}_{\bm d}^H \widetilde{\bm A}_{\bm d} \widetilde{\bm
F}$. Applying the same procedure, we get $\bm \Sigma_x^{-1} =
\widetilde{\bm F}^{-1} \widetilde{\bm \Sigma}_x^{-1} \widetilde{\bm F}$.
The SSE cost for a measurement configuration $\bm d$ then becomes (scaling both terms
with $\lambda$):
\begin{align}
\text{Cost}(\bm{d}) & = \text{tr}\left(\left( {\bm A}_{\bm d}^H {\bm A}_{\bm d} +
\lambda \bm \Sigma_x^{-1} \right)^{-1} \right)
\label{eq:naive_cost}\\
& = \text{tr}\left(\left(\widetilde{\bm F}^{-1} \left( \widetilde{\bm A}_{\bm d}^H
\widetilde{\bm A}_{\bm d} + \lambda \widetilde{\bm \Sigma}_x^{-1} \right)
\widetilde{\bm F}\right)^{-1}\right) \nonumber \\
& = \text{tr}\left(\widetilde{\bm F}^{-1} \left( \widetilde{\bm A}_{\bm d}^H
\widetilde{\bm A}_{\bm d} + \lambda \widetilde{\bm \Sigma}_x^{-1} \right)^{-1}
\widetilde{\bm F}\right) \nonumber \\
& = \text{tr}\left(\left( \widetilde{\bm A}_{\bm d}^H \widetilde{\bm A}_{\bm d} +
\lambda \widetilde{\bm \Sigma}_x^{-1} \right)^{-1} \right)
\label{eq:fast_cost}
\end{align}
where the computational complexity of evaluating \eqref{eq:fast_cost} is much
less than \eqref{eq:naive_cost} due to the diagonalized blocks of
$\widetilde{\bm A}_{\bm d}$ and $\widetilde{\bm \Sigma}_x^{-1}$.

There are two contributors to the complexity of evaluating the cost at
$\text{Cost}(\bm{d})$ for a particular configuration: The multiplication
$\widetilde{\bm A}_{\bm d}^H \widetilde{\bm A}_{\bm d}$, and the inversion
$\left( \widetilde{\bm A}_{\bm d}^H \widetilde{\bm A}_{\bm d} + \lambda
\widetilde{\bm \Sigma}_x^{-1} \right)^{-1}$. In fact, the product
$\widetilde{\bm A}_{\bm d}^H \widetilde{\bm A}_{\bm d}$ only needs to be
calculated once during the algorithm initialization, and it can be efficiently
updated at each iteration by adding/subracting the contribution of the candidate plane
that is iterated over, which can be precomputed.

Thus, the complexity of overall CSBS algorithm is dominated by the inversion
of $\left( \widetilde{\bm A}_{\bm d}^H \widetilde{\bm A}_{\bm d} +
\lambda \widetilde{\bm \Sigma}_x^{-1} \right) \in \mathbb C ^{SN^2 \times SN^2}$
that is performed in each iteration.  While the complexity of a standard inversion algorithm
is $O((SN^2)^3)$, the diagonal structure of this matrix allows for a much faster inversion algorithm
with complexity $O(S^3N^2)$, a speed-up of $N^4$ which is significant for large images.

The total CSBS algorithm complexity is
\vspace{-0.1 in}
$$
O_{sbs} = \sum_{|\bm{d}| = M}^C |\bm{d}| O(S^3N^2) = O(S^3N^2C^2)
$$

\section{Numerical Experiments}
In this section, we present numerical experiments that demonstrate that the
measurement configuration selected by CSBS yields improved reconstructions over
reconstructions obtained from measurements taken at focal planes.
We use a photon sieve as the
diffractive element in our simulations, which offers PSFs with sharper focus
than Fresnel zone plates \cite{kipp2001sharper}.

\begin{figure}[t]
\begin{minipage}[b]{0.32\linewidth}
  \centering
  \centerline{\includegraphics[width=2.5cm]{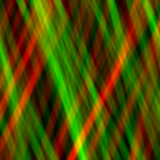}}
  \centerline{(a)}
\end{minipage}
\begin{minipage}[b]{0.32\linewidth}
  \centering
  \centerline{\includegraphics[width=2.5cm]{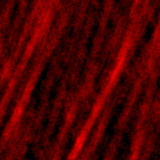}}
  \centerline{(b)}
\end{minipage}
\begin{minipage}[b]{0.32\linewidth}
  \centering
  \centerline{\includegraphics[width=2.5cm]{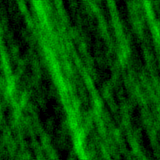}}
  \centerline{(c)}
\end{minipage}

\begin{minipage}[b]{0.32\linewidth}
  \centering
  \centerline{\includegraphics[width=3.6cm]{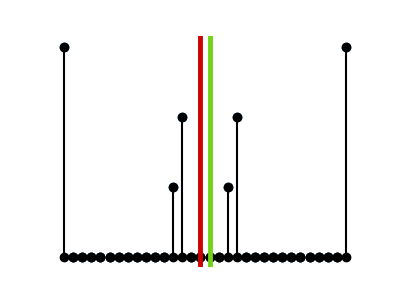}}
  \centerline{(d)}
\end{minipage}
\begin{minipage}[b]{0.32\linewidth}
  \centering
  \centerline{\includegraphics[width=2.5cm]{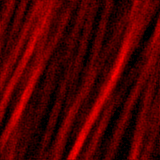}}
  \centerline{(e)}
\end{minipage}
\begin{minipage}[b]{0.32\linewidth}
  \centering
  \centerline{\includegraphics[width=2.5cm]{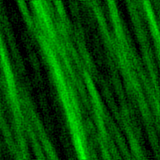}}
  \centerline{(f)}
\end{minipage}
\caption{(a) polychromatic source image with spectral components $\lambda_1$ and
  $\lambda_2$ (b) reconstruction of $\lambda_1$ from focal configuration (c)
  reconstruction of $\lambda_2$ from focal configuration (d) measurement
  locations selected by CSBS (e) reconstruction of
  $\lambda_1$ from CSBS configuration (f) reconstruction of $\lambda_2$ from
  CSBS configuration}
\label{fig:results}
\end{figure}

\vspace{-0.1 in}
We begin by simulating a scenario with two spectral components that are close to
each other in wavelength, shown as separate colors in Figure
\ref{fig:results}(a). We use the MAP estimation framework given in Section 2 as
the image reconstruction algorithm for both the focal plane and CSBS
configurations. For a fair comparison between CSBS and focal plane
reconstructions, we search over $\lambda$ to find the value which maximizes the
focal plane reconstruction \emph{structural similarity} (SSIM) \cite{ssim}, then
use this same $\lambda$ for the CSBS cost function and reconstruction. The final
measurement configuration selected by CSBS is given in Figure
\ref{fig:results}(d), where the two focal planes are marked with red and green
bars. Figures \ref{fig:results}(b) and \ref{fig:results}(c) show the spectral
component reconstructions for the focal plane configuration, and Figures
\ref{fig:results}(e) and \ref{fig:results}(f) for CSBS configuration. The
reconstruction SSIMs for the CSBS and focal plane reconstructions are 0.459
and 0.347, respectively.

\begin{figure}[t]
\begin{minipage}[b]{0.23\linewidth}
  \centering
  \centerline{\includegraphics[width=2cm]{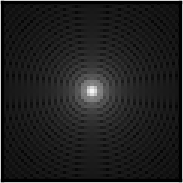}}
  \centerline{(a)}
\end{minipage}
\begin{minipage}[b]{0.23\linewidth}
  \centering
  \centerline{\includegraphics[width=2cm]{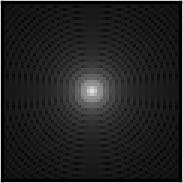}}
  \centerline{(b)}
\end{minipage}
\begin{minipage}[b]{0.23\linewidth}
  \centering
  \centerline{\includegraphics[width=2cm]{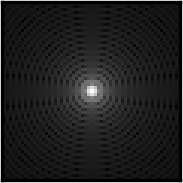}}
  \centerline{(c)}
\end{minipage}
\begin{minipage}[b]{0.23\linewidth}
  \centering
  \centerline{\includegraphics[width=2cm]{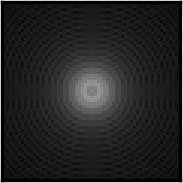}}
  \centerline{(d)}
\end{minipage}
\caption{(a) (b) PSFs at $\lambda_1$ focal plane. (c) (d) PSFs at a measurement location selected by CSBS
  (c) and (d) are less focused that (a) and (b), but have more measurement variation between them.
}
\label{fig:psfs_compare}
\end{figure}

Our intuition on why CSBS chooses out of focus planes pertains to measurement
variation of the PSF pairs for each candidate plane. The spectral components are
very close together in wavelength, so the PSFs corresponding to in focus and out
of focus components at the focal planes are very similar. This leads to poor
measurement variation and makes disentangling the component contributions
difficult. This is especially evident in Figure 3(c), where the features from
one wavelength appear in the reconstruction of the other wavelength. Instead,
CSBS chooses measurement locations where the PSF pairs have more variation at the
expense of a less sharp in focus PSF, shown in Figure \ref{fig:psfs_compare}.

\begin{figure}[htb]
  \begin{minipage}[b]{1\linewidth}
    \centering
    \centerline{\includegraphics[width=8.5cm]{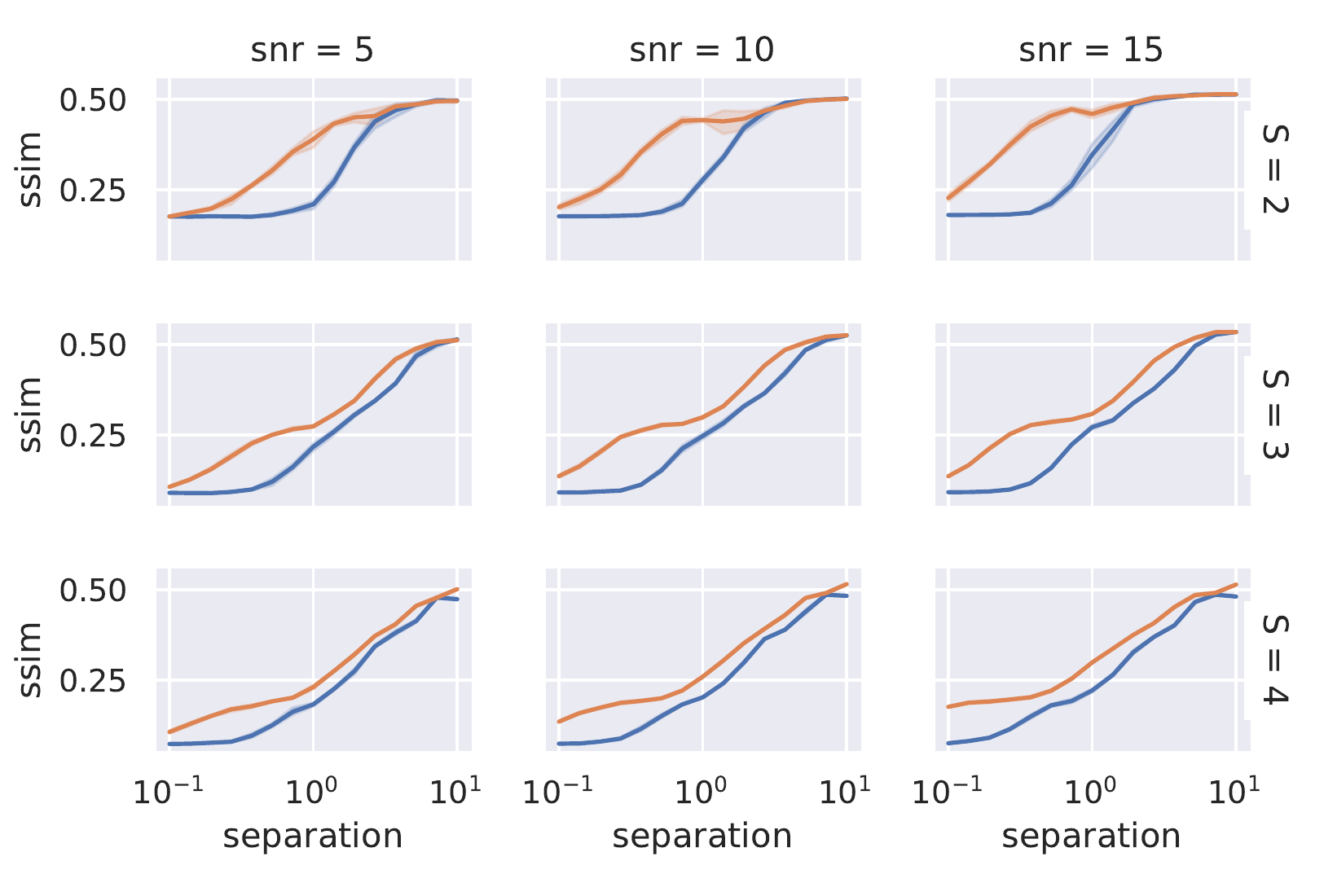}}
  \end{minipage}
  \caption{Reconstruction SSIMs for varied number of spectral components, SNR (dB), and source separation (DOF). CSBS reconstruction SSIM and
  focal plane reconstruction SSIM are shown in orange and blue, respectively.}
  \label{fig:ratio}
\end{figure}

To show that this reconstruction improvement generalizes, we repeat the first experiment for $S = 2, 3, 4$
uniformly spaced spectral components under different noise levels and
spectral component separations measured in depth of focus (DOF) \cite{davila2011high}.  In Figure \ref{fig:ratio}, we plot the mean
SSIM of the reconstructions obtained from measurements at focal planes (blue)
and measurements at planes selected by CSBS (orange).  The CSBS reconstructions
generally have higher SSIM than the focal plane up until the spectral components are sufficiently separated (about 10 DOF), where reconstruction SSIM are about the same.
\vspace{-0.1 in}
\section{Conclusion}
\vspace{-0.1 in}
We apply a variant of the sequential backward selection algorithm to the problem
of acquisition in a diffractive spectral imaging system.  The high
dimensionality of large images makes a direct application of CSBS and SSE cost
computationally intractable, so we have developed a more feasible implementation
of this algorithm and perform an analysis of its complexity to show that it is
significantly faster than the previous implementation for large images.
Finally, we demonstrate CSBS on a simulated spectral imaging system and show that
the optimized measurement configuration achieves equal or better reconstructions than a
choice of measurements at the spectral component focal planes.

\vfill\pagebreak
\clearpage

\bibliographystyle{IEEEbib}
\bibliography{bibliography}

\end{document}